\documentclass[%
preprint,
 amsmath,amssymb,
 aps,
prb,
floatfix,
notitlepage,
]{revtex4-2}

\usepackage{graphicx}
\usepackage{dcolumn}
\usepackage{bm}

\begin{document}

\title{Collective Rayleigh Scattering from Molecular Ensembles under Strong Coupling}

\author{Adina Golombek}
\affiliation{School of Chemistry, Raymond and Beverly Sackler Faculty of Exact Sciences and Tel Aviv University Center for Light-Matter Interaction, Tel Aviv University, Tel Aviv 6997801, Israel.}

\author{Mukundakumar Balasubrahmaniyam}%
\affiliation{School of Chemistry, Raymond and Beverly Sackler Faculty of Exact Sciences and Tel Aviv University Center for Light-Matter Interaction, Tel Aviv University, Tel Aviv 6997801, Israel.}

\author{Maria Kaeek}
\affiliation{School of Chemistry, Raymond and Beverly Sackler Faculty of Exact Sciences and Tel Aviv University Center for Light-Matter Interaction, Tel Aviv University, Tel Aviv 6997801, Israel.}

\author{Keren Hadar}
\affiliation{School of Chemistry, Raymond and Beverly Sackler Faculty of Exact Sciences and Tel Aviv University Center for Light-Matter Interaction, Tel Aviv University, Tel Aviv 6997801, Israel.}

\author{Tal Schwartz}%
\email{Corresponding author \\ talschwartz@tau.ac.il}
\affiliation{School of Chemistry, Raymond and Beverly Sackler Faculty of Exact Sciences and Tel Aviv University Center for Light-Matter Interaction, Tel Aviv University, Tel Aviv 6997801, Israel.}

\begin{abstract}
Rayleigh scattering is usually considered to be the elastic scattering of photons from sub-wavelength physical objects, such as small particles or molecules. Here, we present the spectroscopic study of the scattering properties of molecules embedded in an optical cavity under strong coupling conditions, where the collective interaction between the molecules and the cavity gives rise to composite light-matter excitations known as cavity polaritons. We show that the polaritonic states exhibit strong resonant Rayleigh scattering, reaching \(\sim 25\%\) efficiency. Since the polaritonic wavefunctions in such systems are delocalized, our observations correspond to the collective scattering of each photon from a large ensemble of molecules. 
\end{abstract}

\maketitle


When quantum emitters are placed inside an optical cavity, their interaction with the quantized electromagnetic mode of the cavity may become large enough to overcome the incoherent processes taking place in the system~\cite{Haroche2006}. This regime, which is known as strong light-matter coupling, has been extensively studied in hybrid molecular-photonic systems over the past decade, as it offers exciting possibilities for controlling the photophysical and chemical properties of molecules~\cite{Ebbesen2016,Hertzog2019}. The coupling between an ensemble of molecules and an optical resonator is quantified by the vacuum Rabi frequency, which is roughly given by \(\Omega_R=\frac{2d}{\hbar} \sqrt{\frac{\hbar \omega_c}{2\epsilon V_c}N}\)~\cite{Hertzog2019}. Here $d$ is the transition dipole element of the molecules, \(\hbar\) is Planck's constant, $\omega_c$ is the cavity resonance frequency, \(\epsilon\) is the background dielectric constant inside the cavity, \(V_c\) the cavity mode volume and \(N\) is the number of molecules inside the cavity. The \(\sqrt{N}\) dependence of the coupling strength indicates that, under strong coupling conditions, the interaction between the molecules and the optical mode is collective. As a result, the eigenstates of the coupled system, known as cavity polaritons, represent a coherent superposition of a photon and a material excitation which is delocalized across a macroscopically large ensemble of molecules~\cite{Haroche2006,TAVIS1969}. The collective nature of strong coupling in molecular systems is currently attracting considerable attention~\cite{Sukharev2011,Pino2015,George2016,Vendrell2018,Feist2018,Saez-Blazquez2018}, as it gives rise to fascinating effects. These include, for instance, long-range spatial coherence~\cite{AberraGuebrou2012,Shi2014,Spano2015}, enhanced transport~\cite{Orgiu2015,Schachenmayer2015,Feist2015,Hagenmuller2017,Rozenman2018} and energy transfer~\cite{Coles2014,Zhong2016,Du2018}, and even collective molecular reactivity~\cite{Herrera2016,Galego2017,Groenhof2018,Lather2019}. The progress in this evolving field is intimately linked to the extensive spectroscopic study of organic strongly coupled systems, which has gradually revealed the properties of the polaritonic states and the dynamics of molecules under strong coupling~\cite{Song2004,Virgili2011,Schwartz2013,Canaguier-Durand2015,George2015,Baieva2017,Stranius2018,Martinez-Martinez2018}. Nevertheless, the nature of the polaritonic excitations in organic system as well as the governing dynamical processes, are still not fully understood.

Here, we report on the observation of resonant Rayleigh scattering (RRS) from dye molecules strongly coupled to a planar metallic microcavity and present the spectroscopic study of the scattering from the polaritonic states. Interestingly, our measurements show that the scattering probability depends on the magnitude of the photonic and excitonic components comprising the polaritonic excitations. Considering the hybrid and collective nature of polaritons under strong coupling, in such a process the scattering is fundamentally different from the usual Rayleigh scattering observed close to molecular transitions – under strong coupling, the scattered photons interact coherently with a large ensemble of molecules. In addition, until now, spectroscopic studies of these systems considered either their transmission or (specular) reflection properties. However, as we show here, scattering processes from the polaritonic states are quite significant and therefore cannot be overlooked. We note that in our measurements, we study the elastic scattering from the hybrid cavities, in sharp contrast to previous measurements of Raman scattering in strongly coupled systems~\cite{Tartakovskii,Baieva2013,Nagasawa2014,Shalabney2015}, which is essentially an inelastic process in which coherence is lost.

\begin{figure}[b]
\includegraphics{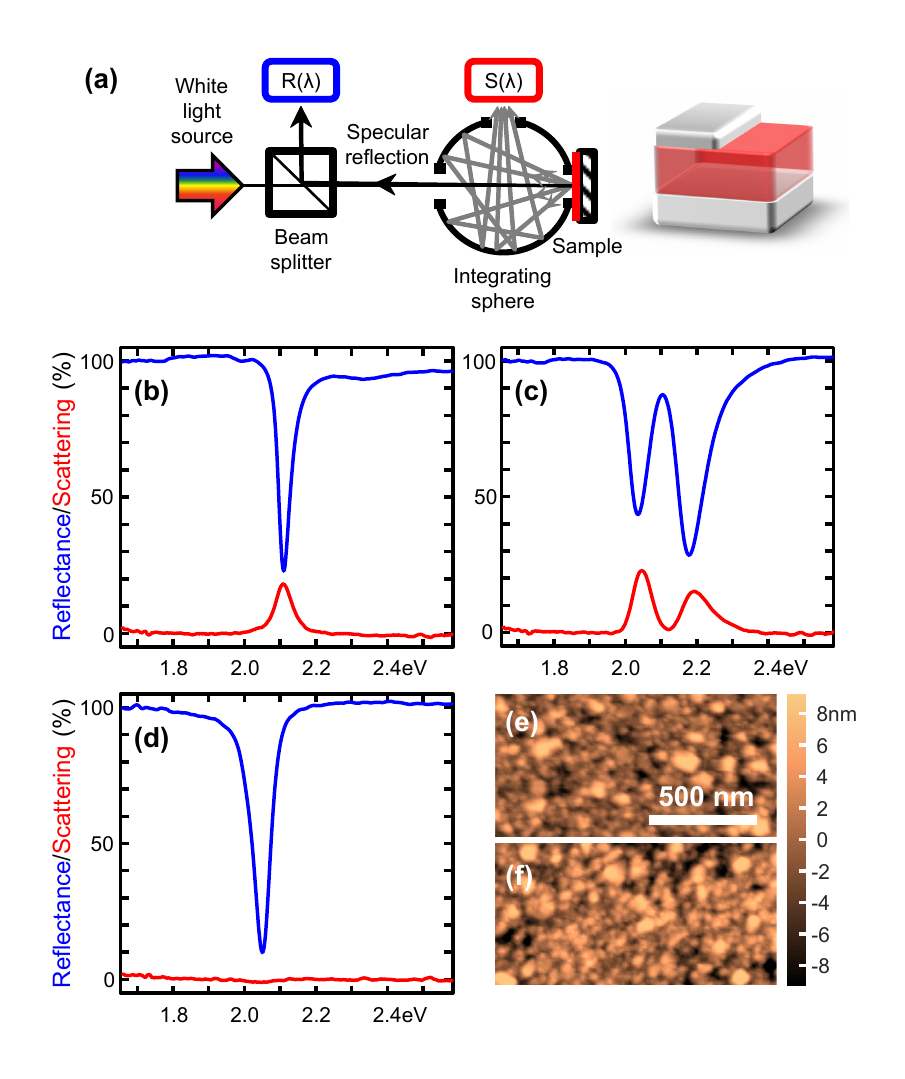}
\caption{\label{fig. 1}
(a) Experimental setup for simultaneous measurement of reflection and scattering using an integrating sphere to collect scattered light. (b), (c) and (d) show the reflectance (blue) and scattering (red) spectra of (b) bare molecules on a silver mirror with a relative concertation of 1:30 mass ratio, (c) the same concentration of molecules embedded in a metallic microcavity tuned to the molecular absorption and (d) a microcavity with no dye molecules. (e) and (f) Atomic force microscope micrographs of an empty cavity (e) and a cavity with TDBC:PVA at a concentration of 1:30 (f).}
\end{figure}

In our experiments, we studied the scattering properties of planar metallic microcavities containing J-aggregates of TDBC molecules~\cite{Schwartz2013} (see Supplementral Material). In order to measure the scattering spectrum, we used a 2" integrating sphere (Thorlabs, IS200-4), as illustrated in Fig. 1(a). As shown, the sample was attached directly to the integrating sphere at its rear output port and a collimated white-light beam was sent though the front port of the integrating sphere onto the sample at normal incidence. The diameter of the beam was limited to \(\sim\)1 mm using an iris, such that a small area of the sample was probed, minimizing the effect of small inhomogeneities in the cavity width. The light scattered from the sample was then collected by the integrating sphere and sent thought the side output port to a fiber-coupled spectrometer. At the same time, the back-reflected beam exiting the integrating sphere was sent to a second fiber-coupled spectrometer, measuring the (specular) reflection spectrum of the sample at exactly the same position. Fig. 1(a) also shows a sketch of the cavity samples used throughout the experiments (see Supplemental Material for further details). The cavity was composed of two Ag mirrors, with the bottom mirror made thick enough (\(\sim\)120 nm) such that no light was transmitted through the sample, and the molecules were embedded in a transparent polymer layer (polyvinyl alcohol, PVA), filling the whole volume between the mirrors. In order to compare the scattering of the cavity-molecules system with the scattering of uncoupled molecules, we deposited the top Ag layer on only one half of the cavity. This ensured that the polymer/TDBC films used in the cavity and non-cavity measurements were identical in terms of thickness and molecular concentration. The reflection signals were normalized to the reflection from a bare, thick Ag layer and the scattering signals were normalized to a highly scattering PTFE reference with a flat spectral response (Thorlabs SM05CP2C). Therefore, our measurements provide absolute quantitative information of the scattering strength.

In Fig. 1(b) we plot the measured reflection (blue line) and scattering (red line) spectra for the bare molecules on the single-mirror side of the sample, with a layer of 1:30 TDBC/PVA mass ratio. As expected, the reflection shows a sharp dip at the absorption maximum of TDBC (2.11 eV). At the same time, a clear scattering peak is observed at the same energy corresponding to resonant Rayleigh scattering from the molecular film~\cite{Ishikawa1997} and displaying 18\% scattering efficiency at the molecular transition energy. On the cavity side of the sample [Fig. 1(c)], the reflection exhibits the characteristic behavior of strong coupling, with the appearance of two reflection dips around the TDBC absorption line (at 2.18 and 2.04 eV), corresponding to the upper and lower polaritons and with a separation of 140 meV between them. Notice that the mean value of these energies (2.11 eV) matches the transition energy of the molecules. This symmetric splitting indicates that the width of the cavity is such that its optical resonance (at normal incidence) is tuned to the molecular transition (see eq. 1 below). The red line in Fig. 1(c) presents the principal result of this study - at the energies of the polaritons we observe two distinct peaks in the scattering spectrum, while the scattering at the bare exciton energy completely disappears. This result shows that under strong coupling, resonant Rayleigh scattering occurs from the polaritonic levels, rather than from the individual, uncoupled molecules. In comparison to a strongly coupled cavity, when performing the same measurements on an empty cavity [containing an undoped PVA layer, Fig. 1(d)], the scattering is lower than 3\%, which is within the noise level, and does not show any particular spectral features. This is despite the fact that a clear reflection dip is observed at the cavity resonance. This shows that the scattering observed for the coupled cavity does not result from surface roughness of the cavity mirrors or some enhancement of scattering due to multiple reflections at the cavity resonance. We note that the lack of scattering from the empty cavity indicates that the observed scattering in the coupled system is fundamentally different than the scattering reported for hybrid molecule-nanoparticle systems~\cite{Zengin2013,Chikkaraddy2016,Tserkezis2018}, in which the optical resonator has a high scattering cross section by itself, even without the molecules. Moreover, AFM measurements of the cavities (Figs. 1e and f) show negligible differences in the roughness of the top mirror between the empty cavity and the cavity containing the molecules (with rms roughness of 3.4 and 4.6 nm, respectively), which cannot explain the strong scattering observed for the cavity-molecules system. We therefore conclude that the scattering peaks shown in Fig. 1(c) can only be associated with the resonant scattering from the molecular ensemble dressed by the cavity electromagnetic mode. We also note that the same kind of resonant scattering from the polaritons is observed when we measure the scattering with a narrow-band excitation while synchronously scanning the scattering wavelength (see Supplemental Material, Fig. S1), rather than measuring the entire spectrum in a single shot. This ensures that the contribution of any emission signal to the measured spectra is negligible, and that the measured spectra truly represent the resonant elastic scattering from the polaritonic states. We note that in previous studies, the absorption of the two polaritonic levels was quantified by measuring the transmission and reflection spectra and then using \(A=1-T-R\) to infer the absorption spectrum~\cite{Schwartz2013,Savona1995,Schwartz2011,Herrera2017}. However, as our measurements indicate scattering from the polaritonic states cannot be neglected and therefore in order to obtain an accurate measure of the overall energy dissipated in the sample one needs to use the relation \(A=1-T-R-S\). Unlike the resonant Rayleigh scattering from polaritonic states that has been observed in strongly coupled systems containing either inorganic semiconductors~\cite{Hayes1998,Freixanet1999,Houdre2000,Shchegrov2000} or organic crystals~\cite{Kena-Cohen2008} and associated to imperfections in the otherwise perfect structure, in our system, the scattering is inherent to the molecular system but modified by the collective nature of strong coupling.

\begin{figure}[b]
\includegraphics{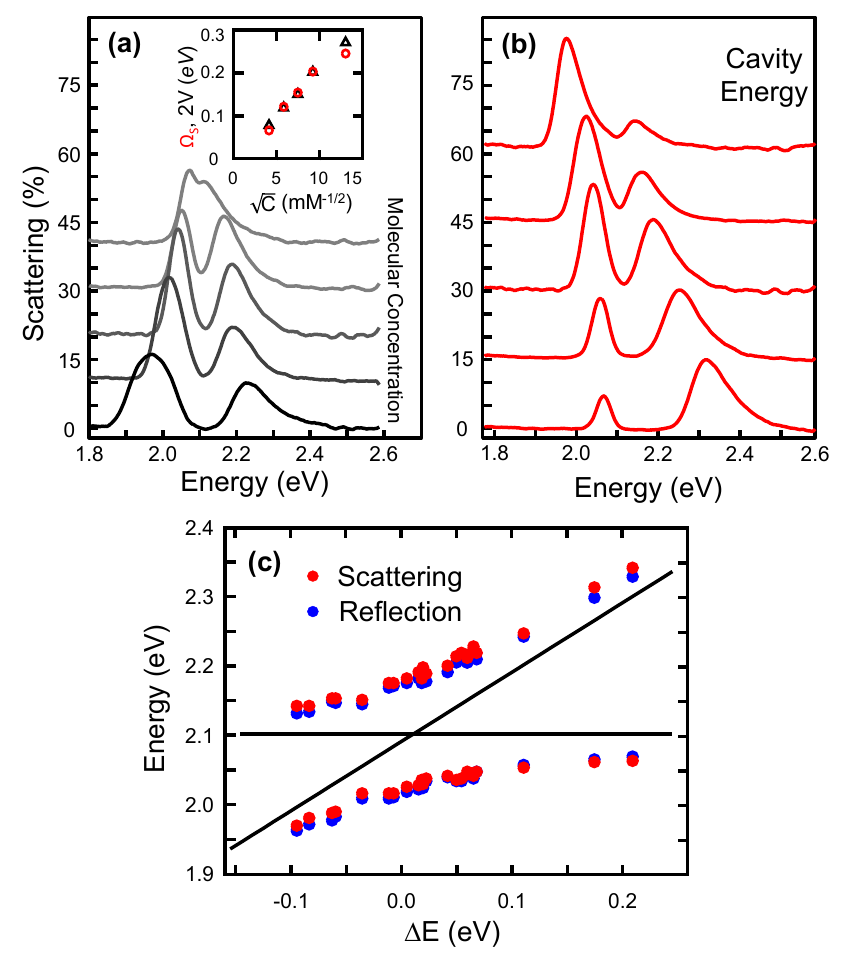}
\caption{\label{fig. 2}
(a) Scattering spectra of cavities at various molecular concentrations. The inset shows the splitting between the scattering peaks ΩS  compared with the coupling strength 2V (extracted from fitting the reflection data to a coupled oscillator model), as a function of the square root of the concentration. (b) Scattering spectra (measured at a concentration of 1:30 TDBC:PVA mass ratio) for various values of tuning. The dashed lines represent fits to the data shown as the sum of a pair of skewed Gaussian functions as described in the text. (c) Dispersion of the strongly coupled cavity at the same molecular concentration as (b), as measured by reflectance (blue) and scattering (red). The dashed lines show the calculated dispersion curves using Eq.~\ref{Eq1}, the solid horizontal line marks the TDBC exciton energy and the diagonal solid line marks the uncoupled photon energy.}
\end{figure}

We repeated the measurements with several different molecular concentrations. Due to the collective coupling of the molecules to the cavity (under strong coupling conditions), the Rabi splitting energy, which is given by 2V, increases as the square root of the concentration~\cite{Ebbesen2016}. Fig. 2(a) shows the measured scattering spectra for TDBC:PVA mass ratios of 1:10, 1:20, 1:30, 1:50 and 1:100, corresponding to molar concentrations of 170, 85, 56, 34 and 17 mM respectively. The splitting in the reflection spectra (not shown) displays the expected square-route dependence on concentration and the polaritonic scattering peaks follow a similar trend [see inset in Fig. 2(a)].

When the cavity resonance is detuned from the molecular absorption line (by varying the thickness of the PVA/TDBC layer), the symmetry of the energy splitting is removed, which is also clearly observed for the scattering peaks, as seen in Fig. 2(b) for several detuning values. The upper (\(E^{(+)})\) and lower (\(E^{(-)})\) polariton energies are given by~\cite{Savona1995}
\begin{equation}
E^{(\pm)}=\frac{1}{2}(E_{c}+E_{x})+\frac{i}{2}\left(\gamma_{c}+\gamma_{x}\right)\pm\sqrt{V^{2}+\frac{1}{4}\left[\Delta-\frac{i}{2}\left(\gamma_{c}-\gamma_{x}\right)\right]^{2}}
\label{Eq1}
\end{equation}
where \(E_c\) is the cavity mode energy, \(E_x=2.11\) eV is the bare exciton energy, \(\Delta=E_c-E_x\) is the detuning between them, \(V\) the coupling strength and \(\gamma_c=60\) meV and \(\gamma_x=40\) meV are the FWHM linewidths of the cavity resonance and the exciton absorption line. In Fig. 2(c) we plot the dispersion curves for the polaritons, acquired from both the reflection dip energies (blue circles) and the scattering peaks (red circles). As can be seen, for all values of detuning the scattering follows the polaritonic dispersion across the entire measured detuning range, up to a slight shift of a few meV. Moreover, by fitting the polariton dispersion curves using Eq.~\ref{Eq1} [dashed lines in Fig. 2(c)] to the data, we extract a coupling strength of 75 meV between the molecules and the cavity.

\begin{figure}[b]
\includegraphics{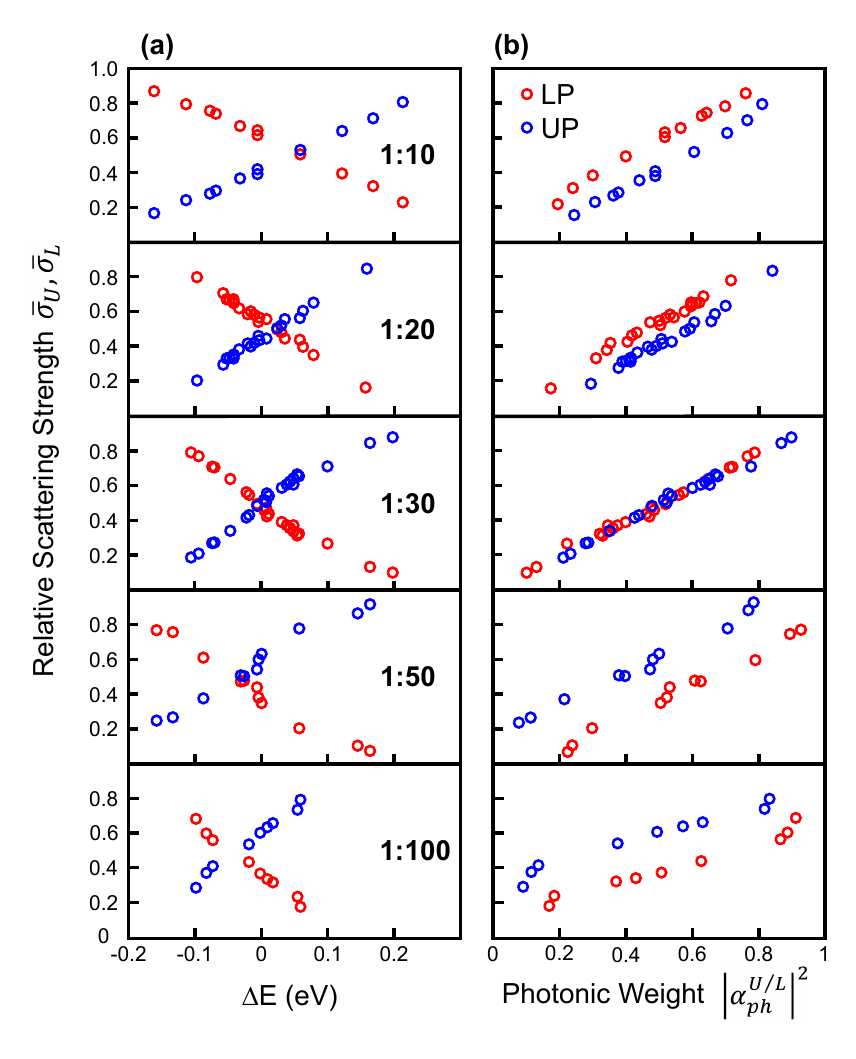}
\caption{\label{fig. 3}
(a) Relative scattering strength of the lower (red) and upper (blue) polaritons as a function of detuning for various molecular concentrations as indicated on the figure. (b) Relative scattering strength of the lower and upper polaritons (colors as above) as a function of their photonic component for various molecular concentrations.}
\end{figure}

Apart from the shift in the scattering peak energies, we also observe [see Fig. 2(b)] that the relative scattering strength of the polaritonic states vary with the detuning. For each one of the scattering spectra we find that the two peaks can be approximated by a pair of skewed Gaussians, in the form \(Ae^{-\left(\frac{E-E_0}{\delta}\right)^2} \times \{1+\text{\textit{erf}}[\frac{\beta}{\sqrt{2}}(\frac{E-E_0}{\delta})]\}\) [dashed lines in Fig. 2(b)]. Using this fit, we calculate the total scattering strength for each one of the polaritons (\(\sigma_U\) and \(\sigma_L\) for the upper and lower polaritons, respectively) by numerical integration. Although the resulting integrated values show quite large fluctuations (see Supplemental Material, Fig. S2) which we attribute to slight sample-to-sample variations in the top mirror thickness, we find that the relative strengths of the polaritons' scattering, given by \(\bar{\sigma}_{U,L}={\sigma_{U,L}}/{(\sigma_U+\sigma_L)}\), exhibit a clear dependence on the detuning. This is shown in Fig. 3(a) for all of the different concentrations measured. As one can see, as the detuning is varied from negative to positive, the scattering from the upper polariton (blue dots) increases while the lower polariton scattering (red dots) decreases. However, for each coupling strength (or molecular concentration), the curves for the relative scattering strengths cross each other at a different non-zero value of the detuning, starting at positive values for large coupling values and going toward negative detuning as the coupling strength becomes smaller.
It is worthwhile remembering that the weights of the polariton constituents (the Hopfield coefficients) also depend on the detuning~\cite{Hertzog2019}. At perfect resonance between the cavity and the molecules, both polaritons are described as an equal mixture of an exciton and a photon, whereas for positive detuning the upper polariton becomes more photonic, while the lower one becomes more excitonic (with an opposite trend for negative detuning). In Fig. 3(b) we plot the relative scattering strengths of the lower and upper polaritons as a function of their photonic component, which is given by
\begin{equation}
    {\left|{\alpha_{ph}^U}\right|}^2=\frac{1}{2}\Bigg(1+\frac{\Delta E}{\sqrt{{\Delta E}^2 + 4 V^2}}\Bigg)
\end{equation}
for the upper polariton and
\begin{equation}
    {\left|{\alpha_{ph}^L}\right|}^2=\frac{1}{2}\Bigg(1-\frac{\Delta E}{\sqrt{{\Delta E}^2 + 4 V^2}}\Bigg)
\end{equation}
for the lower polariton. As seen in Fig. 3(b), the scattering from both polaritons scales linearly with their photonic weight, with a slope of unity, even at relatively large detuning compared to the coupling strength. This trend is quite surprising, since as shown above [Fig. 1(d)] the cavity by itself gives rise to no discernable scattering. The increase in the scattering strength with the photonic weight of the polariton may be related to the collective nature of the polaritons – as the polaritonic states become more photon-like, their wavefunctions become more delocalized~\cite{Michetti}, which may result in larger effective scattering cross sections and higher scattering strength. It is also worthwhile noting that the linear dependences on the photonic weight seen in Fig. 3(b) for the different concentrations exhibit the same slope, irrespective of the coupling strength (notice that at the lowest concentration measured, i.e. 1:100, some deviation from this behavior can be observed. This may directly result from the fact that at this concentration the system is just at the onset of strong coupling, however, it is also possible that the difficulty in fitting the polariton peaks to the coupled oscillator model results in errors for calculated photonic weight of the polaritons and therefore a somewhat different apparent behavior of the scattering strength). When comparing the linear dependences seen in Fig. 3(b), the two lines describing the relation between the scattering strength and the weight of the photonic component for the two polaritons do not necessarily overlap. At high coupling strength the relative strength of the lower polariton is higher than its photonic component, while the strength of upper polariton is correspondingly lower. However, for the samples with smaller coupling strengths this behavior is reversed. This behavior is reminiscent of the asymmetry in the polariton strength intensities observed in semiconductor cavities when the excitonic and photonic widths are varied~\cite{Shchegrov2000}. In our measurements, we observe that the coupling strength also plays a role in determining the relative scattering strengths from the polaritonic states and their variation with the detuning.Specifically at perfect resonance, where the coupled-oscillator model predicts identical properties for the two polaritonic states, a clear symmetry breaking between them is observed. Such symmetry breaking is known to occur for incoherent relaxation processes in such systems~\cite{Virgili2011,Schwartz2013,George2016,Herrera2017,Neuman2018}. Interestingly, as seen by our measurements, this also exists in the short time-scale ellastic scattering processes.

In conclusion, we have shown that the resonant Rayleigh scattering from dye molecules embedded in a cavity is strongly affected by the collective strong coupling between the molecules and the cavity mode. As a result, the elastic scattering of photons does not occur from the individual molecules themselves, but rather from the polaritonic states, which encompass a large ensemble of molecules. Our spectroscopic measurements show that the magnitude of the scattering increases linearly with the weight of the photonic component of the polaritons, even though the scattering from an empty cavity is  negligible and does not display any resonant behavior. The observed resonant scattering from the strongly coupled molecules reveals yet another aspect of the spectroscopic properties of the collective light-matter excitations formed in such systems. These results, together with further investigation of the scattering process such as its angular distribution~\cite{Freixanet1999,Houdre2000,Kena-Cohen2008} or ultrafast temporal dynamics~\cite{Hayes1998,Freixanet1999} may lead to a better understanding of polaritons in organic systems, their spatially extended structure and the role of disorder in such strongly coupled systems~\cite{Michetti,Haakh2016}.

\bibliography{manuscript}

\end{document}